\documentclass[conference]{IEEEtran}
\IEEEoverridecommandlockouts
\usepackage{cite}
\usepackage{amsmath,amssymb,amsfonts}
\usepackage{graphicx}
\usepackage{textcomp}
\usepackage{xcolor}
\usepackage{array}
\usepackage{tabularx}
\usepackage{booktabs}
\usepackage{algorithm}
\usepackage{algpseudocode}
\def\BibTeX{{\rm B\kern-.05em{\sc i\kern-.025em b}\kern-.08em
    T\kern-.1667em\lower.7ex\hbox{E}\kern-.125emX}}
\begin{document}

\title{Real-time Pipe Burst Localization in Water Distribution Networks Using Change Point Detection Algorithms}

\author{\IEEEauthorblockN{Takudzwa Mzembegwa}
\IEEEauthorblockA{\textit{Dept. of Computer Science} \\
\textit{University of the Western Cape}\\
Cape Town, South Africa \\
3805515@myuwc.ac.za}
\and
\IEEEauthorblockN{Clement N Nyirenda}
\IEEEauthorblockA{\textit{Dept. of Computer Science} \\
\textit{University of the Western Cape}\\
Cape Town, South Africa  \\
0000-0002-4181-0478}
}

\maketitle

\begin{abstract}
Change point detection (CPD) has proved to be an effective tool for detecting drifts in data and its use over the years has become more pronounced due to the vast amount of data and IoT devices readily available. This study analyzes the effectiveness of Cumulative Sum (CUSUM) and Shewhart Control Charts for identifying the occurrence of abrupt pressure changes for pipe burst localization in Water Distribution Network (WDN). Change point detection algorithms could be useful for identifying the nodes that register the earliest and most drastic pressure changes with the aim of detecting pipe bursts in real-time. TSNet, a Python package, is employed in order to simulate pipe bursts in a WDN. The pressure readings are served to the pipe burst localization algorithm the moment they are available for real-time pie burst localization. The performance of the pipe burst localization algorithm is evaluated using a key metric such as localization accuracy under different settings to compare its performance when paired with either CUSUM or Shewhart.  Results show that the pipe burst localization algorithm has an overall better performance when paired with CUSUM. Although, it does show great accuracy for both CPD algorithms when pressure readings are being continuously made available without a big gap between time steps. The proposed approach however still needs further experiments on different  WDNs to assess the performance and accuracy of the algorithm on real-world WDN models.
\end{abstract}

\section{Introduction}
Pipe bursts in WDNs cause the loss of a great deal of treated water. If not located on time, they not only lead to huge water loss, but they also end up causing WDN and environmental contamination, and they also lead to great financial losses to water service providers. To address this challenge, a lot of research work has been put into improving pipe burst detection and localization. Recent work mainly focuses on data-driven approaches as data can now be acquired in real-time with relative ease through the use of data acquisition systems like the supervisory control and data acquisition (SCADA) system. With good quality data, data-driven approaches making use of machine learning algorithms have proved to be effective at detecting and locating pipe bursts offline. Multi-Layer Perceptron (MLP) have been employed for pipe burst localization this involved training an artificial neural network (ANN) with simulations of a real network in Portugal over 24 hours \cite{capelo2021near}. Support Vector Regression and Deep learning have also been applied to water leakage detection \cite{kemba2017leakage}, \cite{mzembegwa2023comparison}, \cite{zhou2019deep}. In \cite{mzembegwa2023comparison}, deep learning in a form of a Fully linear ResNet model was employed and it had a mean accuracy of 98.23\% for pipe burst localization with zero false positives \cite{mzembegwa2023comparison}. The model required 24 hours' worth of data to locate the burst pipe. There has been some work done to reduce the required wait time before locating a pipe burst. Srirangarajan et al \cite{srirangarajan2013wavelet} proposed a wavelet-based method for burst detection, along with a graph-based localization technique. The approach was validated on the WaterWiSe@SG wireless sensor network test bed over 2 hours \cite{srirangarajan2013wavelet}. Although the approaches provide good results their long wait times before they can locate a pipe burst lead to a considerable loss of treated water and allow the pipe bursts to become an environmental hazard.

The use of machine learning to locate pipe bursts in real-time in the real world suffers because existing WDN infrastructure lacks tools to capture nodal data like pressure and flow, resulting in a scarcity of extensive historical pressure data. Even when historical pressure data is available, it may not be labelled, or there may be insufficient historical pipe burst data to efficiently train machine learning models for accurate and reliable pipe burst localization. Additionally, machine learning approaches often require a substantial amount of data to be captured after a pipe burst occurs before they can infer the burst locations \cite{zhang2023real}. Longer wait times before a pipe burst is located lead to an increase in Non-Revenue Water (NRW) as water service providers' response times will be slower. It is, therefore, important to devise an efficient real-time pipe burst localization technique. This will allow burst pipes to be identified as soon as a pipe burst is detected. Existing research on pipe burst localization, while successful in some cases, heavily relies on substantial historical data and takes time to locate burst pipes after they occur. However, these approaches often fail to consider the lack of real-world historical data with sufficient information about previous pipe bursts. This highlights the imperative need for pipe burst localization techniques that do not depend on historical data.

Research on real-time pipe burst localization in WDNs has received considerable attention. However, many existing methods have significant wait times, often ranging from minutes to hours. Notable real-time pipe burst detection has been achieved by Zhang et al. \cite{zhang2023real} for gas pipelines, and by Gupta et al. \cite{gupta2021pipeline} for locating bursts in a single 50-meter pipeline and three T-shaped PVC pipelines. Zhang et al \cite{zhang2023real} proposed an attention mechanism-based long short-term memory (AM-LSTM) approach for real-time gas pipeline leak detection and localization, the localization process involved identifying the sensor closest to the leak. Validated through lab experiments, their method achieved detection and localization of leaks at five positions, by outputting higher weights at sensors near the leaks. Gupta et al \cite{gupta2021pipeline}, introduced a method for detecting and localizing pipeline bursts using pressure transient analysis. They used wavelet de-noising and CUSUM analysis to identify bursts from sensor-collected pressure signals. Validated on a single 50-meter and 3 T-shaped PVC pipelines, their method effectively detected bursts and reduced localization errors to within 3 meters, showing potential for real-time detection using transient analysis of pressure signals \cite{gupta2021pipeline}. Comparing predicted and observed pressure values in WDNs is an approach that is also commonly employed for real-time pipe burst detection \cite{zaman2020review}, \cite{zhang2022real}, \cite{soldevila2016leak}. Some of these approaches can be referred to as near-real-time as seen in the work of Zhang et al \cite{zhang2022real}, who proposed a near real-time a pipe burst detection method that triggers 15 minutes after a burst event. The method combines predicted deviation, pressure variation, and absolute pressure at multiple time steps, utilizing Western Electric Company criteria, a decision tree model, and an SVR model for pipe burst detection. Such methods are however affected by inaccuracies in prediction methods comparing predicted and observed pressure values, and the availability of historical data to train the prediction models. These examples highlight the need to develop approaches that can locate pipe bursts in WDNs in real-time without the need of historical data that is rarely ever available in real-world WDN. 

In this paper, we propose a real-time pipe burst localization approach for WDNs that can locate pipe bursts in as little as one second on a WDN with multiple links. This approach makes use of CPD to detect transient pressure changes, amplitude and ultimately the localization of a pipe burst using the transient information. This paper also explores the simulation of burst events using TSNet for use in real-time pipe burst localization and the effectiveness of CPD algorithms in locating pipe bursts. Two algorithms are discussed: CUSUM and Shewhart Control Charts. For each algorithm, a brief description is provided. 

The rest of the paper is organized as follows. Section II presents the concept of Water Distribution Networks (WDNs). Section III presents the change point detection mechanisms that have been used in this work, while Section IV presents the real-time pipe burst localization approach taken. Section V then presents the results and discussion. Finally, section VI concludes the paper.

\section{Water distribution networks}
A Water Distribution Network (WDN) is a vital component of civil infrastructure that supplies fresh water for residential consumption, industrial activities, and firefighting \cite{bui2020water}. A typical WDN comprises various components, including junctions, pipes, valves, pumps, tanks, and reservoirs \cite{sangroula2022optimization}, all working together to form a complex system aimed at distributing water for these uses.

\subsection{Water Distribution Network Hydraulic Model}
Change point detection algorithms are dependent on data; in this study, real-time time series data, composed of simulated pressure values, is used. Therefore, the pipe burst simulation process is extremely important, as the consistency and quality of the simulated pressure readings will influence the effectiveness of the algorithms and the usability of the results in real-world WDNs. In this work, EPANET 2.2 is used to build the WDN hydraulic model and export it as an INP file for transient simulations. The governing equations for hydraulic transients are nonlinear hyperbolic Partial Differential Equations (PDE), so a closed-form solution is not available \cite{lee2022hydraulic}. Numerical methods must be used to solve these governing equations \cite{lee2022hydraulic}. There are several numerical methods such as, but not limited to, Methods of Characteristics (MOC), implicit Finite Difference (FD) method, and explicit FD \cite{lee2022hydraulic}. Therefore, a Python package Transient Simulations in Water Networks (TSNet) is then employed to import the INP file and perform extended-period transient simulations. TSNet utilizes the Method of Characteristics (MOC) to solve the governing equations for transient flow.

\subsection{Transient simulations}
Transient refers to a temporary or intermediate state that occurs during a transition between two steady states in a system \cite{xu2017overview}. In WDNs, transients can arise from abrupt alterations in pump or valve configurations, the starting or stopping of pump operations and the occurrences of pipe bursts \cite{xu2017overview}. Transient events can unveil a great amount of meaningful data about the WDN as transient waves swiftly travel through the pipelines \cite{rossman2000epanet}. Consequently, transient models remain essential for simulating and extrapolating flow conditions across the entire system by utilizing data gathered from several sensors. To this end, the pressure data in this work is generated from a TSNet, a Python package that allows for the simulation of transient system responses. TSNet was created to provide researchers with an open-source package for transient simulations to cater for the ever-growing interest in transient-based approaches \cite{xing2020transient}. The package simulates bursts, leaks, valve closure, surge tank, and pump shut-off \cite{xing2020transient}. It can be integrated with other case-specific applications, e.g., sensor placement, event detection, model calibration, and sensitivity analysis \cite{xing2020transient}. Before conducting a transient simulation, it is essential to establish initial steady-state conditions, encompassing pipe flows and nodal heads. To achieve this, TSNet makes use of Water Network Tool for Resilience WNTR \cite{klise2017water}, a python package, TSNet is built on top of, to simulate the network's steady state to set up the starting conditions for subsequent transient simulations \cite{xing2020transient}. WNTR facilitates the incorporation of both demand-driven and pressure-driven hydraulic scenarios, while also accommodating the introduction of background leaks. This capability enables the execution of pipe burst transient simulations on a network that includes background leaks. Pressure pressure-driven demand (PDD) simulations are employed as they provide a more accurate depiction of pressure loss during pipe burst scenarios \cite{mzembegwa2023comparison}. The package makes use of MOC to solve the governing equations for transient flow \cite{xing2020transient}. During simulations, TSNet treats demands as pressure-driven discharges. Consequently, the actual demands will deviate from those initially defined by the user. The actual demands are modelled by utilizing the immediate pressure head at the node in conjunction with demand discharge coefficients \cite{xing2020transient}. The pressure-driven demand approach enables dynamic fluctuations in actual demands in response to localized pressure variations, yielding a portrayal of more authentic conditions \cite{xing2020transient} \cite{kanakoudis2016assessing}.

\section{Changepoint Detection Methods}
Thus section presents the two changepoint detection (CPD) algorithms that have been used in this work, namely CUSUM and Shewhart.

\subsection{CUSUM}
The CUSUM algorithm was initially introduced by Page \cite{page1954continuous}, and it has since become a prominent algorithm in change point detection and statistical quality control. The CUSUM algorithm functions by monitoring the cumulative sum of discrepancies between successive data points. It is particularly effective in identifying abrupt or gradual alterations within a process. In this study, the CUSUM algorithm was applied to track two cumulative sums, \( C_{i-1}^+ \) and \( C_{i-1}^- \), which respectively represent the cumulative positive and negative deviations between consecutive data points which are defined by
\begin{equation}
C_{i}^+ = max(0, C_{i-1}^+ + x_{i} - x_{i-1} - drift)\label{eq1}
\end{equation}
and
\begin{equation}
C_{i}^- = max(0, C_{i-1}^- - x_{i} - x_{i-1} - drift)\label{eq2}
\end{equation}
where \( x_{i}\) and \( x_{i-1}\) are the current and previous data points, and drift represents a predefined bias. The drift correction is there in order to prevent the algorithm from detecting a change when there isn't a genuine alteration or when there's a gradual shift. The cumulative sums are reset to zero whenever they become negative. Change points are detected when the cumulative sums \( C_{i-1}^+ \) and \( C_{i-1}^- \) exceed a predetermined threshold. This indicates that a significant change has occurred in the time series data. An alarm is then triggered, and the algorithm records the alarm index and start index and resets the cumulative sums to zero for subsequent monitoring. The code snippet in algorithm \ref{alg:cusum_code} illustrates the functionality of CUSUM.

\begin{algorithm}
\caption{CUSUM Algorithm for Pipe Burst Detection\cite{xing2020transient}} \label{alg:cusum_code}
\begin{algorithmic}[1]
\Function{cusum}{time, x, thr, dr, ending}
    \State $x \gets \text{atleast1d}(x)$, $time \gets \text{atleast1d}(time)$
    \State $gp, gn, gp\_real, gn\_real \gets \text{zeros}(x.size)$
    \State $ta, s, e \gets \text{array}([], [], [])$
    \State $tap, tan \gets 0, 0$
    \State $amp \gets \text{array}()$
    
    \For{$i \gets 1$ to $x.size$}
        \State $s \gets x[i] - x[i-1]$, $gp[i] \gets gp[i-1] + s - dr$
        \State $gp\_real[i] \gets gp\_real[i-1] + s$, $gn[i] \gets gn[i-1] - s - dr$
        \State $gn\_real[i] \gets gn\_real[i-1] - s$
        
        \If{$gp[i] < 0$} \State $gp[i], gp\_real[i], tap \gets 0, 0, i$ \EndIf
        \If{$gn[i] < 0$} \State $gn[i], gn\_real[i], tan \gets 0, 0, i$ \EndIf
        
        \If{$gp\_real[i] > thr$ or $gn\_real[i] > thr$}
            \State $ta \gets \text{append}(ta, i)$
            \If{$gp\_real[i] > thr$} \State $s \gets \text{append}(s, tap)$ \Else \State $s \gets \text{append}(s, tan)$ \EndIf
            \State $gp[i], gn[i] \gets 0, 0$, $gp\_real[i], gn\_real[i] \gets 0, 0$
        \EndIf
    \EndFor

    \If{$s.size > 0$}
        \State $s2, \_, \_ \gets \text{cusum}(time[::-1], x[::-1], thr, dr)$
        \State $e \gets x.size - s2[::-1] - 1$
        \State $s, j \gets \text{unique}(s, \text{return\_index=True})$
        \State $ta \gets ta[j]$ 
        
        \If{$s.size != e.size$}
            \If{$s.size < e.size$}
                \State $e \gets e[[\text{max}(e >= i) \text{ for } i \text{ in } ta]]$
            \Else
                \State $j \gets [\text{max}(i >= ta[::-1])-1 \text{ for } i \text{ in } e]$
                \State $ta, s \gets ta[j], s[j]$
            \EndIf
        \EndIf
        
        \State $j \gets e[:-1] - s[1:] > 0$
        \If{$j.any()$}
            \State $ta \gets ta[~\text{append}(False, j)]$
            \State $s \gets s[~\text{append}(False, j)]$
            \State $e \gets e[~\text{append}(j, False)]$
        \EndIf
        \State $amp \gets x[e] - x[s]$
    \EndIf
    \State \Return $s, e, amp$
\EndFunction
\end{algorithmic}
\end{algorithm}

\subsection{Shewhart control charts}
Shewhart control charts were first introduced by Shewhart \cite{shewhart1930economic}. The Shewhart control chart method employs a moving average and control limits to identify changes in data distribution. These charts are firmly grounded in statistical principles, as they are closely linked to traditional statistical hypothesis testing \cite{koutras2007statistical}. The fundamental concept of a Shewhart control chart algorithm is to detect a point in time when there is a shift in the parameters of a distribution using control limits. The control limits, as defined in \cite{nist2012shewhart}, are established as:
\begin{equation}
UCL = \mu + s \times  \sigma; LCL = \mu - s \times  \sigma;\label{eq3}
\end{equation}
where \( \mu\) and \( \sigma\) are the window mean and window standard deviation respectively. And \(s\) is a constant multiplier that determines how wide the control limits are. Data outside the upper control limit (\(UCL\)) and lower control limit (\(LCL\)) are indicative of potential shifts or an out-of-control process. Change points are detected when a data point surpasses the established \(UCL\) or goes below the established \(LCL\). Such an occurrence signifies a deviation from the expected distribution, prompting the algorithm to record the anomaly's index. The code snippet in algorithm \ref{alg:cusum_shewhart} illustrates the functionality of Shewhart Control Chart.

\begin{algorithm}
\caption{Shewhart Algorithm for Pipe Burst Detection}  \label{alg:cusum_shewhart}
\begin{algorithmic}[1]
\Function{detectShewhart}{time, x, threshold}
\State \textbf{Input:}
\State \hspace{\algorithmicindent} time : 1D array \Comment{time}
\State \hspace{\algorithmicindent} x : 1D array \Comment{data}
\State \hspace{\algorithmicindent} threshold : \Comment{amplitude threshold}

\State \textbf{Output:}
\State \hspace{\algorithmicindent} tai : 1D array \Comment{index of when the change started}
\State \hspace{\algorithmicindent} amp : 1D array \Comment{amplitude of changes}
\State $x \gets \text{atleast1d}(x)$
\State $time \gets \text{atleast1d}(time)$
\State $tai \gets \text{array}()$
\State $amp \gets \text{array}()$
\State $mean \gets \text{mean}(x)$
\State $std \gets \text{std}(x)$
\For{$i \gets 1$ to $x.size$}
    \If{$|x[i] - mean| > threshold * std$ and $|x[i] - mean| > threshold$}
        \State $tai \gets \text{append}(tai, i)$
    \EndIf
\EndFor
\State $amp \gets |x[tai] - mean|$
\State \Return $tai, amp$
\EndFunction
\end{algorithmic}
\end{algorithm}

\section{Real-time Pipe Burst Localization}
Pipe bursts, defined as sudden pipe rupture and break events, can introduce sudden and rapid hydraulic transients, which then propagate in the pipe system \cite{lee2022hydraulic}. In TSNet, pipe bursts can be specified at junctions. To model the burst occurring along a pipe, the user should introduce a new junction at the location of the burst in the INP file \cite{lee2022hydraulic}.

\subsection{Real-time Pipe Burst Simulation}
To run the simulation and distribute the nodal pressure from the WDN during pipe bursts, a number of steps are taken, as follows:
\paragraph{Import the WDN} After the WDN hydraulic model was created in EPANET 2.0. The INP file of the model is imported to create a transient model with TSNet.
\paragraph{Define Network Characteristics} On top of the information already defined in the INP file, we set the wave speed and time step and the duration of the simulation. TSNet set the wave speed to 1200m/s by default. The time step is not defined as TSNet will adjust it to simultaneously satisfy two constraints, each pipe must satisfy the Courant's criterion and all pipes in the network must have the same timestep. TSNet will use the maximum allowed time step for cases where the time step is not defined \cite{xing2020transient}.
\paragraph{Create a Burst Event} To create a burst event we utilize WNTR to split a pipe with the burst into 2 separate pipes connected by a new node. A burst event is then added to the new node. The new node thus represents the pipe burst orifice as water is ejected into the atmosphere at the new node.
\paragraph{Initialize Transient Model} Before initializing the transient model a few values have to be set, that is, the time the simulation starts and the engine used for the steady-state simulation. With the values set to 0 sec and PDD for the time the simulation starts and the steady-state engine, respectively. Once the values are set the model is then initialized. That is, WNTR is run to simulate the network's steady state, establishing initial conditions that serve as a foundation for subsequent transient simulations.
\paragraph{Execute Transient Simulation} Execute the transient simulation using the MOCSimulator. TSNet employs the Method of Characteristics to solve the governing transient flow equations. Once the simulation is complete it returns a Python object containing the simulation results, including time series data, like node pressure, flow rate, node burst discharge, and velocity and it includes the simulation timestamps.
\paragraph{Real-time pressure transmission} Real-time pressure transmission: At each time step, the nodal pressure from each pressure meter is immediately transmitted to the central system. When a pressure meter captures a reading, it is promptly made available to the central system for that time step. This continuous, real-time data provision persists until the simulation is complete.

\subsection{Pipe burst localization}
To identify the location of the pipe burst a number of steps are taken.
\paragraph{Network graph} To get the structure of the WDN we create a directed acyclic graph. TSNet has an inbuilt function from WNTR, to\verb|_|graph, to convert a transient model of a WDN to a NetworkX graph. NetworkX is a Python library for studying graphs and networks. The function is called against the model and is supplied with link\verb|_|weight and modify\verb|_|direction. Link\verb|_|weight is a \(dataframe\) with the link weights of the graph. In this study, the link weights are used to determine the direction of the links. Therefore they are all set to either \(1\) or \(-1\). The link weights come from pipe flow rates that are converted to 1 if positive and -1 when negative. The parameter modify\verb|_|direction set to \(true\), this allow the links with a negative link weight to have their start node and end node switched, and the absolute link weight assigned to the link.
\paragraph{Fetching data} Read the data that is being made available by the pressure meters as it come along. Data is read after \(interval\) new complete pressure readings have been made available. The system keeps on attempting to read the data until the \(interval\) new complete pressure readings are read and then the algorithm continues. When the data is available the system concatenates the previous \(interval\) pressure readings with the current \(interval\) pressure reading. A time step has complete pressure readings if every pressure meter has made its reading for the time step available.
\paragraph{Detecting nodal pressure changes} For each junction node with a pressure meter, run the chosen CPD algorithm. If the node has a drastic pressure change for the current \(interval\), store the current node, change start times and amplitude of each change.
\paragraph{Identify node linked to the burst source} With the results from the CPD algorithm we then find the node that registered the biggest pressure change. That is, finding the node that has the absolute maximum amplitude of all the nodes that registered a drastic pressure change. The first node with the biggest pressure change in the transient state is most likely linked to the burst pipe. Pressure waves caused by the pipe burst will have to propagate throughout the network, therefore it is going to take more time for the pressure waves to reach nodes that are further from the burst source. Thus, the node with the biggest pressure drop (\(node\)) in the transient state is linked to the burst source.
\paragraph{Burst pipe localization} The \(node\) identified in the previous step is then used as a starting point for finding the burst pipe. Using the WDN directed acyclic graph, we find the predecessors and neighbours of the \(node\). For instances where the CPD algorithm only registered 2 nodes with drastic pressure changes, we check if one of the neighbours of \(node\) registered drastic pressure changes. If it did, we conclude that the link between \(node\) and its neighbour is the burst source. Otherwise, we check if \(node\) only has one predecessor. If it does, we conclude that the bust source is between the predecessor and \(node\). If the burst source is still not found, we find the predecessor of \(node\) that has the biggest absolute amplitude given that the predecessor registered a drastic pressure change. Otherwise, we find the nodal means of the concatenation of the previous \(interval\) pressure readings with the current \(interval\) pressure readings. We then find the predecessor of \(node\) that has the lowest average pressure and conclude that the burst source is between the predecessor and \(node\).

\section{Results and discussion}
\begin{figure}[htbp]
\centerline{\includegraphics[width=\linewidth]{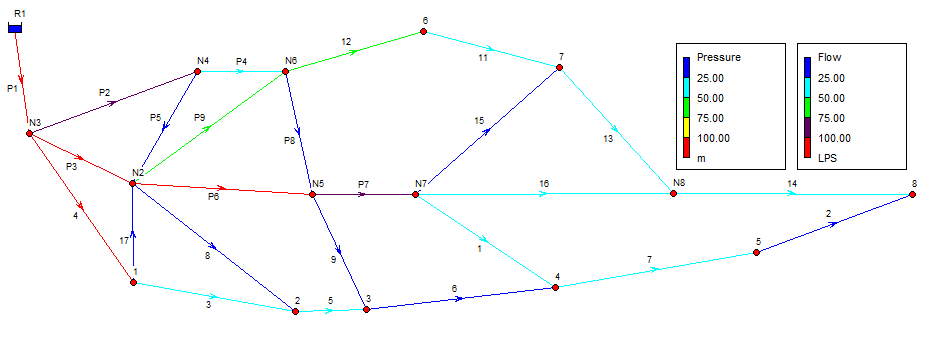}}
\caption{WDN Hydraulic model in Epanet.}
\label{fig}
\end{figure}

To validate the proposed pipe burst localization method, a hydraulic model of a WDN was used. The WDN consisted of, a reservoir, 15 junction nodes and 25 pipes Fig \ref{fig}. 
The WDN was used to simulate pipe bursts in a real-time manner for all the different capture time intervals used in this study. A capture time interval is the time interval at which the simulation is supposed to capture a pressure reading and make it available for pipe burst localization after the previous pressure reading (e.g., a capture time interval of 0.2 seconds means the pressure meters are capturing and transmitting WDN pressure readings every 0.2 seconds). The capture time intervals experimented with in this study were 0.2 seconds, 2 seconds, 5 seconds and 10 seconds, as shown in Tables \ref{tab:cusum_scenarios} and \ref{tab:shewhart_scenarios}.

The thresholds and localization interval of both the CUSUM and Shewhart were calculated via a trial and error approach to maximise the number of correct pipe bursts located. The 4 capture intervals were selected in an attempt to find the capture interval that produced the best results. The localization intervals were kept the same for both CUSUM and Shewhart to make the results from both approaches comparable. 

% Define a new column type for centered text with fixed width
\newcolumntype{Y}{>{\centering\arraybackslash}p{2cm}}
\begin{table}[htbp]
    \caption{Parameters used for CUSUM the approach}
    \centering
        \begin{tabular}{|c|Y|Y|Y|}
            \hline
            Scenario & Capture Interval (sec) & Threshold (m) & Localization interval \\
            \hline
            S\verb|_|C1 & 0.2 & 0.3 & 5 \\
            \hline
            S\verb|_|C2 & 2 & 2 & 10 \\
            \hline
            S\verb|_|C3 & 5 & 10 & 10 \\
            \hline
            S\verb|_|C4 & 10 & 13 & 10 \\
            \hline
        \end{tabular}
    \label{tab:cusum_scenarios}
\end{table}

\begin{table}[htbp]
    \caption{Parameters used for the Shewhart approach}
    \centering
        \begin{tabular}{|c|Y|Y|Y|}
            \hline
            Scenario & Capture Interval (sec) & Threshold (m) & Localization interval \\
            \hline
            S\verb|_|S1 & 0.2 & 1.5 & 5 \\
            \hline
            S\verb|_|S2 & 2 & 3 & 10 \\
            \hline
            S\verb|_|S3 & 5 & 2 & 10 \\
            \hline
            S\verb|_|S4 & 10 & 3 & 10 \\
            \hline
        \end{tabular}
    \label{tab:shewhart_scenarios}
\end{table}

The Real-time Pipe Burst Localization method was run independently for both the CUSUM and Shewhart approach using the simulated real-time data. The algorithm was tested by adding one pipe burst at a time at each pipe for all pipes in the WDN. The result of each localization attempt was captured and is shown in Tables \ref{tab:cusum_results} and \ref{tab:shewhart_results}. The results show the start and end node of the pipe that presumably has a pipe burst. A result \(R1 - N3\), means the burst pipe is the pipe is between \(R1\) and \(N3\), with \(R1\) being the start node and \(N3\) the end node. The flow direction is from the start node to the end node, that is \(R1\) to \(N3\). The pipe bursts, for the different scenarios, that were correctly located are shown in bold with the total of the correctly identified pipe bursts shown at the bottom of the individual tables.

% Define a new column type for centered text with fixed width
\newcolumntype{C}{>{\centering\arraybackslash}p{1.4cm}}
\begin{table}[ht]
    \caption{Burst link localization results using CUSUM}
    \centering
        \begin{tabular}{|c|C|C|C|C|}
            \hline
            Pipe & CUSUM (S\verb|_|C1) & CUSUM (S\verb|_|C2) & CUSUM (S\verb|_|C3) & CUSUM (S\verb|_|C4) \\
            \hline
            P1 & \textbf{R1 - N3} & N7 - 7 & 7 - N8 & \textbf{R1 - N3} \\
            \hline
            P2 & \textbf{N3 - N4} & \textbf{N3 - N4} & \textbf{N3 - N4} & \textbf{N3 - N4} \\
            \hline
            P3 & \textbf{N3 - N2} & N8 - 8 & N2 - 2 & N4 - N2 \\
            \hline
            P4 & \textbf{N4 - N6} & 4 - 5 & \textbf{N4 - N6} & \textbf{N4 - N6} \\
            \hline
            P5 & \textbf{N4 - N2} & N7 - N8 & N3 - N4 & N3 - N4 \\
            \hline
            P6 & \textbf{N2 - N5} & N5 - N7 & N6 - N5 & N6 - N5 \\
            \hline
            P7 & N6 - N5 & \textbf{N5 - N7} & N8 - 8 & \textbf{N5 - N7} \\
            \hline
            P8 & \textbf{N6 - N5} & N7 - N8 & N2 - N6 & N2 - N6 \\
            \hline
            P9 & \textbf{N2 - N6} & N6 - 6 & \textbf{N2 - N6} & \textbf{N2 - N6} \\
            \hline
            1 & \textbf{N7 - 4} & \textbf{N7 - 4} & 4 - 5 & 5 - 8 \\
            \hline
            2 & \textbf{5 - 8} & \textbf{5 - 8} & 4 - 5 & 4 - 5 \\
            \hline
            3 & \textbf{1 - 2} & \textbf{1 - 2} & \textbf{1 - 2} & \textbf{1 - 2} \\
            \hline
            4 & \textbf{N3 - 1} & \textbf{N3 - 1} & \textbf{N3 - 1} & \textbf{N3 - 1} \\
            \hline
            5 & \textbf{2 - 3} & \textbf{2 - 3} & \textbf{2 - 3} & \textbf{2 - 3} \\
            \hline
            6 & \textbf{3 - 4} & \textbf{3 - 4} & 4 - 5 & 4 - 5 \\
            \hline
            7 & 3 - 4 & 4 - 5 & \textbf{4 - 5} & \textbf{4 - 5} \\
            \hline
            8 & \textbf{N2 - 2} & \textbf{N2 - 2} & \textbf{N2 - 2} & \textbf{N2 - 2} \\
            \hline
            9 & \textbf{N5 - 3} & \textbf{N5 - 3} & \textbf{N5 - 3} & 2 - 3 \\
            \hline
            11 & \textbf{6 - 7} & \textbf{6 - 7} & \textbf{6 - 7} & \textbf{6 - 7} \\
            \hline
            12 & \textbf{N6 - 6} & \textbf{N6 - 6} & \textbf{N6 - 6} & \textbf{N6 - 6} \\
            \hline
            13 & \textbf{7 - N8} & \textbf{7 - N8} & \textbf{7 - N8} & \textbf{7 - N8} \\
            \hline
            14 & \textbf{N8 - 8} & \textbf{N8 - 8} & 5 - 8 & 5 - 8 \\
            \hline
            15 & \textbf{N7 - 7} & \textbf{N7 - 7} & \textbf{N7 - 7} & \textbf{N7 - 7} \\
            \hline
            16 & \textbf{N7 - N8} & \textbf{N7 - N8} & 7 - N8 & 7 - N8 \\
            \hline
            17 & N3 - 1 & 1 - N2 & N3 - 1 & N3 - 1 \\
            \hline
            \hline
            \textbf{Correct:} & \underline{\textbf{23}} & \underline{\textbf{18}} & \underline{\textbf{13}} & \underline{\textbf{16}} \\
            \hline
        \end{tabular}
    \label{tab:cusum_results}
\end{table}

\begin{table}[ht]
    \caption{Burst link localization results using Shewhart}
    \centering
     \begin{tabular}{|c|C|C|C|C|}
        \hline
        Pipe & Shewhart (S\verb|_|S1) & Shewhart (S\verb|_|S2) & Shewhart (S\verb|_|S3) & Shewhart (S\verb|_|S4) \\
        \hline
        P1 & \textbf{R1 - N3} & 4 - 5 & N5 - 3 & N3 - N4 \\
        \hline
        P2 & \textbf{N3 - N4} & \textbf{N3 - N4} & N8 - 8 & \textbf{N3 - N4} \\
        \hline
        P3 & \textbf{N3 - N2} & 5 - 8 & N2 - 2 & N4 - N2 \\
        \hline
        P4 & N3 - N4 & 4 - 5 & N6 - 6 & N6 - N5 \\
        \hline
        P5 & \textbf{N4 - N2} & 4 - 5 & N8 - 8 & N3 - N4 \\
        \hline
        P6 & \textbf{N2 - N5} & 5 - 8 & N5 - N7 & N6 - N5 \\
        \hline
        P7 & \textbf{N5 - N7} & \textbf{N5 - N7} & N8 - 8 & \textbf{N5 - N7} \\
        \hline
        P8 & \textbf{N6 - N5} & N8 - 8 & N6 - 6 & N2 - N6 \\
        \hline
        P9 & \textbf{N2 - N6} & N6 - 6 & N5 - N7 & \textbf{N2 - N6} \\
        \hline
        1 & \textbf{N7 - 4} & \textbf{N7 - 4} & 4 - 5 & 3 - 4 \\
        \hline
        2 & \textbf{5 - 8} & \textbf{5 - 8} & 4 - 5 & 4 - 5 \\
        \hline
        3 & \textbf{1 - 2} & \textbf{1 - 2} & \textbf{1 - 2} & \textbf{1 - 2} \\
        \hline
        4 & \textbf{N3 - 1} & \textbf{N3 - 1} & 2 - 3 & \textbf{N3 - 1} \\
        \hline
        5 & \textbf{2 - 3} & \textbf{2 - 3} & \textbf{2 - 3} & \textbf{2 - 3} \\
        \hline
        6 & \textbf{3 - 4} & \textbf{3 - 4} & 4 - 5 & \textbf{3 - 4} \\
        \hline
        7 & \textbf{4 - 5} & \textbf{4 - 5} & \textbf{4 - 5} & \textbf{4 - 5} \\
        \hline
        8 & \textbf{N2 - 2} & \textbf{N2 - 2} & \textbf{N2 - 2} & \textbf{N2 - 2} \\
        \hline
        9 & \textbf{N5 - 3} & \textbf{N5 - 3} & \textbf{N5 - 3} & 2 - 3 \\
        \hline
        11 & \textbf{6 - 7} & \textbf{6 - 7} & \textbf{6 - 7} & \textbf{6 - 7} \\
        \hline
        12 & \textbf{N6 - 6} & \textbf{N6 - 6} & 7 - N8 & \textbf{N6 - 6} \\
        \hline
        13 & \textbf{7 - N8} & \textbf{7 - N8} & \textbf{7 - N8} & \textbf{7 - N8} \\
        \hline
        14 & \textbf{N8 - 8} & \textbf{N8 - 8} & 5 - 8 & 5 - 8 \\
        \hline
        15 & \textbf{N7 - 7} & \textbf{N7 - 7} & \textbf{N7 - 7} & \textbf{N7 - 7} \\
        \hline
        16 & \textbf{N7 - N8} & \textbf{N7 - N8} & 7 - N8 & \textbf{N7 - N8} \\
        \hline
        17 & \textbf{1 - N2} & \textbf{1 - N2} & N3 - 1 & N3 - 1 \\
        \hline
        \hline
        \textbf{Correct:} & \underline{\textbf{24}} & \underline{\textbf{18}} & \underline{\textbf{8}} & \underline{\textbf{15}} \\
        \hline
    \end{tabular}
    \label{tab:shewhart_results}
\end{table}

The first scenario with a capture interval of 0.2 seconds produced the best results for both CUSUM and Shewhart. This could be because of the frequency at which the localization system is receiving nodal pressure data. A smaller time interval means that the smallest pressure changes that occur in a short period are captured. Thus, a smaller localization interval is rich in information as it allows for the capturing of the smallest pressure deviations that happen in the earliest part of the transient state before the WDN starts to transition to a steady state. These readings in the earliest part of the transient state carry important information as these pressure readings can reflect the difference in time at which the nodes registered pressure changes. This is because nodes closest to the burst source are likely to reflect the pressure changes first before the pressure waves propagate to the rest of the WDN.

For scenarios 2, 3 and 4, a localization interval of 10 was used. A localization interval of 10 was big enough for the algorithms to locate a pipe burst without drastically increasing the time it takes to locate a pipe burst after it has occurred, although it did in seconds. Smaller localization intervals did not provide good results in the trial-and-error run. This could be because a smaller capture interval does not contain a lot of information about the early stages of the transient state, so it requires a big localization window to identify significant changes within the pressure data. 

Scenario 2, with a capture interval of 2 had the better performance between scenarios 2, 3 and 4. This can be explained by the fact that 2 seconds still contains relatively more information about the early stages of the of a transient state as pressure readings are still relatively captured in a short enough window that can reflect the slower arrival of pressure waves at other nodes. Between scenarios 3 and 4, scenario 4 had the better performance, with scenario 3 showing the worst performance of all the scenarios.  This can be due to the capture interval being too big to capture crucial information that is present in the early stages of the transient state. A capture interval of 10 misses crucial information in the early stages of a transient state but is long enough to reflect the pressure changes on a longer time scale.

\begin{table}[htbp]
    \caption{Comparison of CUSUM and Shewhart`s accuracy}
    \centering
        \begin{tabular}{|c|Y|Y|}
            \hline
            Capture interval & CUSUM & Shewhart \\
            \hline
            0.2 & 92\% & \textbf{96\%} \\
            \hline
            2 & \textbf{72\%} & \textbf{72\%} \\
            \hline
            5 & \textbf{52\%} & 32\%  \\
            \hline
            10 & \textbf{64\%} & 60\% \\
            \hline
            \hline
            \textbf{Average} & \underline{\textbf{70\%}} & \underline{\textbf{65\%}} \\
            \hline
        \end{tabular}
    \label{tab:cusum_shewhart_comp}
\end{table}

The localization algorithm proved to work well with both CPD algorithms when the capture interval was 0.2 seconds. The two algorithms were compared by their accuracy which was calculated as follows:
\begin{equation}
Acc = \dfrac{P_c}{P_t}\times100\%\label{eq4}
\end{equation}
Where \(P_c\) and\( P_t\) are the total number of correctly located pipes and the total number of pipes in the water distribution network. Both algorithms had an accuracy of more than 90\% on a 0.2 seconds capture interval. Shewhart however had a better performance of 96\% when using a 0.2 second capture interval compared to CUSUM`s 92\%. For this study, Shewhart was modified to reduce its sensitivity which then reduced the number of detected false positives. A pressure change has to fall outside the \(UCL\) and \(LCL\), and the pressure change also has to be greater than a stated threshold for the pressure change to be considered as caused by a pipe burst. This change improved the performance of the algorithm. For scenario 2, the two algorithms had the same accuracy of 72\% and the pipe burst they successfully located were the same. The pipes that were not correctly located are linked to end nodes that are being supplied water by multiple pipes except for \(P1\) which has the reservoir as a start node. Therefore, the reason they might have not been located correctly is because a capture interval missed crucial pressure changes that happen on a millisecond level before the other pipes start supplying the end node with more water to rebalance the WDN pressure. For scenarios 3 and 4, CUSUM had a better performance of 52\% and 64\%, respectively, compared to Shewhart`s 32\% to 60\%. CUSUM`s average performance of 70\% over all the scenarios was better than the 65\% from Shewhart. Shewhart was outperformed by CUSUM, particularly in instances where the capture interval was larger. This can be explained by the nature of the algorithm, which utilizes the mean and standard deviation. The deviation between pressure data increases significantly if the intervals at which the pressure data is captured are greatly separated, as WDN pressure drops drastically during the transient state. This larger deviation results in wider control limits, making the algorithm less sensitive to detecting smaller pressure changes, and thus, incorrectly locating pipe bursts\cite{george2007economic}. CUSUM was only outperformed in scenario 1.

\section{Conclusion}

Transient analysis has proved to be an effective method for pipe burst detection and localization. In this paper, real-time transient data was utilised in proposing a novel real-time pipe burst localization approach, which can identify the pipe that is a source of a pipe burst. The proposed approach made use of CPD algorithms to identify drastic nodal pressure changes in the transient state which was in turn used to find the start and end nodes of a burst pipe. CUSUM proved to have a better overall performance of 70\%, compared to Shewhart`s 65\% pipe burst localization accuracy. A capture interval of 0.2 seconds with a localization interval of 5, proved to be the most accurate. Scoring 96\% accuracy for Shewhart and 92\% for CUSUM. That is, the algorithm could accurately locate a pipe burst a second after it has occurred.

Although the proposed approach proved to be accurate and fast at locating pipe bursts, further experiments need to be carried out on different WDNs to assess the performance and accuracy of the algorithm on more complex WDN models and real-world WDN data. Parameter optimization approaches for the proposed approach also need to be explored in order to improve the accuracy and performance of the model.

\bibliographystyle{IEEEtran} % Specify the bibliography style
%\bibliography{mybib} % Specify the path to your .bib file

\begin{thebibliography}{10}
\providecommand{\url}[1]{#1}
\csname url@samestyle\endcsname
\providecommand{\newblock}{\relax}
\providecommand{\bibinfo}[2]{#2}
\providecommand{\BIBentrySTDinterwordspacing}{\spaceskip=0pt\relax}
\providecommand{\BIBentryALTinterwordstretchfactor}{4}
\providecommand{\BIBentryALTinterwordspacing}{\spaceskip=\fontdimen2\font plus
\BIBentryALTinterwordstretchfactor\fontdimen3\font minus \fontdimen4\font\relax}
\providecommand{\BIBforeignlanguage}[2]{{%
\expandafter\ifx\csname l@#1\endcsname\relax
\typeout{** WARNING: IEEEtran.bst: No hyphenation pattern has been}%
\typeout{** loaded for the language `#1'. Using the pattern for}%
\typeout{** the default language instead.}%
\else
\language=\csname l@#1\endcsname
\fi
#2}}
\providecommand{\BIBdecl}{\relax}
\BIBdecl

\bibitem{capelo2021near}
M.~Capelo, B.~Brentan, L.~Monteiro, and D.~Covas, ``Near--real time burst location and sizing in water distribution systems using artificial neural networks,'' \emph{Water}, vol.~13, no.~13, p. 1841, 2021.

\bibitem{kemba2017leakage}
J.~Kemba, K.~Gideon, and C.~N. Nyirenda, ``Leakage detection in tsumeb east water distribution network using epanet and support vector regression,'' in \emph{2017 IST-Africa Week Conference (IST-Africa)}.\hskip 1em plus 0.5em minus 0.4em\relax IEEE, 2017, pp. 1--8.

\bibitem{mzembegwa2023comparison}
T.~S. Mzembegwa and C.~Nyirenda, ``A comparison of fully-linear deep learning methods for pipe burst localization in water distribution networks,'' in \emph{2023 IST-Africa Conference (IST-Africa)}.\hskip 1em plus 0.5em minus 0.4em\relax IEEE, 2023, pp. 1--8.

\bibitem{zhou2019deep}
X.~Zhou, Z.~Tang, W.~Xu, F.~Meng, X.~Chu, K.~Xin, and G.~Fu, ``Deep learning identifies accurate burst locations in water distribution networks,'' \emph{Water research}, vol. 166, p. 115058, 2019.

\bibitem{srirangarajan2013wavelet}
S.~Srirangarajan, M.~Allen, A.~Preis, M.~Iqbal, H.~B. Lim, and A.~J. Whittle, ``Wavelet-based burst event detection and localization in water distribution systems,'' \emph{Journal of Signal Processing Systems}, vol.~72, pp. 1--16, 2013.

\bibitem{zhang2023real}
X.~Zhang, J.~Shi, M.~Yang, X.~Huang, A.~S. Usmani, G.~Chen, J.~Fu, J.~Huang, and J.~Li, ``Real-time pipeline leak detection and localization using an attention-based lstm approach,'' \emph{Process Safety and Environmental Protection}, vol. 174, pp. 460--472, 2023.

\bibitem{gupta2021pipeline}
A.~Gupta and K.~Kulat, ``Pipeline burst detection and its localization using pressure transient analysis,'' in \emph{Proceedings of the International Conference on Paradigms of Computing, Communication and Data Sciences: PCCDS 2020}.\hskip 1em plus 0.5em minus 0.4em\relax Springer, 2021, pp. 13--25.

\bibitem{zaman2020review}
D.~Zaman, M.~K. Tiwari, A.~K. Gupta, and D.~Sen, ``A review of leakage detection strategies for pressurised pipeline in steady-state,'' \emph{Engineering Failure Analysis}, vol. 109, p. 104264, 2020.

\bibitem{zhang2022real}
X.~Zhang, Z.~Long, T.~Yao, H.~Zhou, T.~Yu, and Y.~Zhou, ``Real-time burst detection based on multiple features of pressure data,'' \emph{Water Supply}, vol.~22, no.~2, pp. 1474--1491, 2022.

\bibitem{soldevila2016leak}
A.~Soldevila, J.~Blesa, S.~Tornil-Sin, E.~Duviella, R.~M. Fernandez-Canti, and V.~Puig, ``Leak localization in water distribution networks using a mixed model-based/data-driven approach,'' \emph{Control Engineering Practice}, vol.~55, pp. 162--173, 2016.

\bibitem{bui2020water}
K.~Bui, M.~Marlim, and D.~Kang, ``Water network partitioning into district metered areas: a state-of-the-art review. water 12 (4), 1002,'' 2020.

\bibitem{sangroula2022optimization}
U.~Sangroula, K.-H. Han, K.-M. Koo, K.~Gnawali, and K.-T. Yum, ``Optimization of water distribution networks using genetic algorithm based sop--wdn program,'' \emph{Water}, vol.~14, no.~6, p. 851, 2022.

\bibitem{lee2022hydraulic}
\BIBentryALTinterwordspacing
J.~Lee, L.~Xing, and L.~Sela, ``{Hydraulic transients in pipe systems},'' in \emph{{Embracing Analytics in the Drinking Water Industry}}.\hskip 1em plus 0.5em minus 0.4em\relax IWA Publishing, 06 2022. [Online]. Available: \url{https://doi.org/10.2166/9781789062380\_0215}
\BIBentrySTDinterwordspacing

\bibitem{xu2017overview}
X.~Xu and B.~Karney, ``An overview of transient fault detection techniques,'' \emph{Modeling and monitoring of pipelines and networks: Advanced tools for automatic monitoring and supervision of pipelines}, pp. 13--37, 2017.

\bibitem{rossman2000epanet}
L.~A. Rossman \emph{et~al.}, ``0,'' \emph{Water Supply and Water Resources Division, National Risk Management Research Laboratory, Cincinnati, OH}, vol. 45268, 2000.

\bibitem{xing2020transient}
L.~Xing and L.~Sela, ``Transient simulations in water distribution networks: Tsnet python package,'' \emph{Advances in Engineering Software}, vol. 149, p. 102884, 2020.

\bibitem{klise2017water}
K.~A. Klise, D.~Hart, D.~M. Moriarty, M.~L. Bynum, R.~Murray, J.~Burkhardt, and T.~Haxton, ``Water network tool for resilience (wntr) user manual,'' Sandia National Lab.(SNL-NM), Albuquerque, NM (United States), Tech. Rep., 2017.

\bibitem{kanakoudis2016assessing}
V.~Kanakoudis and K.~Gonelas, ``Assessing the results of a virtual pressure management project applied in kos town water distribution network,'' \emph{Desalination and Water Treatment}, vol.~57, no.~25, pp. 11\,472--11\,483, 2016.

\bibitem{page1954continuous}
E.~S. Page, ``Continuous inspection schemes,'' \emph{Biometrika}, vol.~41, no. 1/2, pp. 100--115, 1954.

\bibitem{shewhart1930economic}
W.~A. Shewhart, ``Economic quality control of manufactured product 1,'' \emph{Bell System Technical Journal}, vol.~9, no.~2, pp. 364--389, 1930.

\bibitem{koutras2007statistical}
M.~Koutras, S.~Bersimis, and P.~Maravelakis, ``Statistical process control using shewhart control charts with supplementary runs rules,'' \emph{Methodology and Computing in Applied Probability}, vol.~9, pp. 207--224, 2007.

\bibitem{nist2012shewhart}
\BIBentryALTinterwordspacing
``Shewhart control chart,'' Apr 2012. [Online]. Available: \url{https://doi.org/10.18434/M32189}
\BIBentrySTDinterwordspacing

\bibitem{george2007economic}
\BIBentryALTinterwordspacing
G.~Nenes and G.~Tagaras, ``An economic comparison of cusum and shewhart charts,'' \emph{IIE Transactions}, vol.~40, no.~2, pp. 133--146, 2007. [Online]. Available: \url{https://doi.org/10.1080/07408170701592499}
\BIBentrySTDinterwordspacing

\end{thebibliography}

\end{document}